
\newskip\oneline \oneline=1em plus.3em minus.3em
\newskip\halfline \halfline=.5em plus .15em minus.15em
\newbox\sect
\newcount\eq
\newbox\lett
\newdimen\short
\def\adv{\global\advance\eq by1}
\def\set#1#2{\setbox#1=\hbox{#2}}
\def\nextlet#1{\global\advance\eq by-1\setbox
                \lett=\hbox{\rlap#1\phantom{a}}}

\newcount\eqncount
\newcount\sectcount
\eqncount=0
\sectcount=0
\def\sectadv{\global\advance\sectcount by1    }
\def\secta{\global\advance\sectcount by1    }
\def\equn{\global\advance\eqncount by1\eqno{(\copy\sect.\the\eqncount)} }
\def\put#1{\global\edef#1{(\the\sectcount.\the\eqncount)}           }

\magnification = 1200{\rm}
\voffset 0.2 truein
\vsize   8.7 truein
\hsize  6.25 truein
\hfuzz 30 pt

\def\mbox#1#2{\vcenter{\hrule \hbox{\vrule height#2in
                \kern#1in \vrule} \hrule}}  
\def\sq{\,\raise.5pt\hbox{$\mbox{.09}{.09}$}\,}
\def\sqb{\,\raise.5pt\hbox{$\overline{\mbox{.09}{.09}}$}\,}
\def\tri{\triangle}

\def\nabar{{\overline \nabla}}
\def\bR{\overline R}
\def\R{R^2}
\def\Ric{R_{ab}R^{ab}}
\def\Rie{R_{abcd}R^{abcd}}
\def\sqR{\sq R}

 2

\def\a{\alpha}
\def\b{\beta}

\def\d{\delta}
\def\e{\epsilon}

\def\g{\gamma}

\def\k{\kappa}
\def\l{\lambda}

\def\o{\omega}
\def\p{\pi}

\def\r{\rho}
\def\s{\sigma}

\def\D{\Delta}

\def\G{\Gamma}

\def\O{\Omega}

\def\cc{{\cal C}}

\def\ck{{\cal K}}

\def\cm{{\cal M}}

\def\co{{\cal O}}

\def\cy{{\cal Y}}

\def\pa{\partial}
\def\de{\nabla}

\def\TH{{\raise.2ex\hbox{$\displaystyle \bigodot$}\mskip-4.7mu \llap H \;}}
\def\face{{\raise.2ex\hbox{$\displaystyle \bigodot$}\mskip-2.2mu \llap {$\ddot
        \smile$}}}

\def\Hat#1{\rlap{\kern.10em$\widehat{\phantom G}$}#1}
\def\HAt#1{\rlap{\kern.05em$\widehat{\phantom G}$}#1}

\def\cap#1{\rlap{\kern.1em$\widehat{\phantom{G\vrule height.8em}}$}#1{}}
\def\Cap#1{\rlap{\kern.05em$\widehat{\phantom{G\vrule height.8em}}$}#1{}}

\def\leftrightarrowfill{$\mathsurround=0pt \mathord\leftarrow \mkern-6mu
        \cleaders\hbox{$\mkern-2mu \mathord- \mkern-2mu$}\hfill
        \mkern-6mu \mathord\rightarrow$}
\def\overleftrightarrow#1{\vbox{\ialign{##\crcr
        \leftrightarrowfill\crcr\noalign{\kern-1pt\nointerlineskip}
        $\hfil\displaystyle{#1}\hfil$\crcr}}}

\def\frac#1#2{{\textstyle{#1\over\vphantom2\smash{\raise.20ex
        \hbox{$\scriptstyle{#2}$}}}}}
\def\ha{\frac12}

\catcode`@=11
\def\underline#1{\relax\ifmmode\@@underline#1\else
        $\@@underline{\hbox{#1}}$\relax\fi}
\catcode`@=12

\def\nis{\nointerlineskip}
\def\Abar{\vbox{\nis\moveright.33em\vbox{
        \hrule width.35em height.04em}\nis\kern.05em\hbox{$A$}}{}}
\def\Dbar{\vbox{\nis\moveright.20em\vbox{
        \hrule width.50em height.04em}\nis\kern.05em\hbox{$D$}}{}}
\def\Gbar{\vbox{\nis\moveright.20em\vbox{
        \hrule width.50em height.04em}\nis\kern.05em\hbox{$G$}}{}}
\def\mbar{\vbox{\nis\moveright.15em\vbox{
        \hrule width.60em height.04em}\nis\kern.05em\hbox{$m$}}{}}
\def\Rbar{\vbox{\nis\moveright.20em\vbox{
        \hrule width.50em height.04em}\nis\kern.05em\hbox{$R$}}{}}
\def\Vbar{\vbox{\nis\moveright.05em\vbox{
        \hrule width.60em height.04em}\nis\kern.05em\hbox{$V$}}{}}
\def\Xbar{\vbox{\nis\moveright.20em\vbox{
        \hrule width.60em height.04em}\nis\kern.05em\hbox{$X$}}{}}
\def\thetabar{\vbox{\nis\moveright.15em\vbox{
        \hrule width.30em height.04em}\nis\kern.05em\hbox{$\theta$}}{}}
\def\Lambdabar{\vbox{\nis\moveright.25em\vbox{
        \hrule width.35em height.04em}\nis\kern.05em\hbox{${\mit\Lambda}$}}{}}
\def\Sigmabar{\vbox{\nis\moveright.25em\vbox{
        \hrule width.50em height.04em}\nis\kern.05em\hbox{${\mit\Sigma}$}}{}}
\def\phibar{\vbox{\nis\moveright.18em\vbox{
        \hrule width.40em height.04em}\nis\kern.05em\hbox{$\phi$}}{}}
\def\chibar{\vbox{\nis\moveright.12em\vbox{
        \hrule width.40em height.04em}\nis\kern.05em\hbox{$\chi$}}{}}
\def\psibar{\vbox{\nis\moveright.23em\vbox{
        \hrule width.40em height.04em}\nis\kern.05em\hbox{$\psi$}}{}}
\def\debar{\vbox{\nis\moveright.18em\vbox{
        \hrule width.35em height.04em}\nis\kern.05em\hbox{$\partial$}}{}}
\def\delbar{\vbox{\nis\moveright.10em\vbox{
        \hrule width.63em height.04em}\nis\kern.05em\hbox{$\nabla$}}{}}

\def\rarr{\rightarrow}

\def\Y{Y_{_{JM}}}
\def\Ys{Y^*_{_{JM}}}

\def\dT{{\dot{T}}_{JM}^{(+)}}

\def\jo{J_1}
\def\jt{J_2}
\def\mo{M_1}
\def\mt{M_2}


\def\cAntMot{1}
\def\cAMM{2}
\def\cPoly{3}
\def\cdeS{4}
\def\cKPZ{5}
\def\cYork{6}
\def\cpapI{7}
\def\cLiou{8}
\def\cLBC{9}
\def\cGSW{10}
\def\cSch{11}

\nopagenumbers
\rightline{CPTH-S377.0995}
\rightline{LA-UR-95-3420}
\rightline{hep-th/9509169}
\rightline{September 24, 1995}
\vskip 1.2truecm
\centerline{\bf {PHYSICAL STATES OF}}
\centerline{\bf {THE QUANTUM CONFORMAL FACTOR}}
\vskip 1.2truecm
\centerline{Ignatios Antoniadis}
\centerline{{\it Centre de Physique Th\'eorique
}\footnote{$^{\dag}$}{\it Laboratoire Propre du CNRS UPR A.0014}}
\centerline{{\it Ecole Polytechnique}}
\centerline{{\it 91128 Palaiseau, France}}
\vskip .5truecm
\centerline{Pawel O. Mazur}
\centerline{{\it Dept. of Physics and Astronomy}}
\centerline{{\it University of South Carolina}}
\centerline{{\it Columbia, SC 29208 USA}}
\vskip .3truecm
\centerline{and}
\vskip .3truecm
\centerline{Emil Mottola}
\centerline{{\it Theoretical Division, T-8}}
\centerline{{\it Mail Stop B285}}
\centerline{{\it Los Alamos National Laboratory}}
\centerline{{\it Los Alamos, NM 87545 USA}}
\vskip 1truecm \centerline{\bf Abstract}
\vskip .5truecm
The conformal factor of the spacetime metric becomes dynamical due to the
trace anomaly of matter fields. Its dynamics is described by an effective
action which we quantize by canonical methods on the Einstein universe
$R\times S^3$. We find an infinite tower of discrete states which satisfy
the constraints of quantum diffeomorphism invariance. These physical states
are in one-to-one correspondence with operators constructed by integrating
integer powers of the Ricci scalar.

\vfill
\eject

\baselineskip=20pt
\footline={\hss\tenrm\folio\hss}
\pageno = 1

\newcount\eqncount
\sectadv
\set\sect{1}
\beginsection{1. Introduction}

It is virtually certain that at the ultrashort Planck scale  a theory of
gravitational interactions requires a framework quite different from the
familiar classical metric description of spacetime. At larger distance scales
the metric becomes a useful variable and gravity should be described by an
effective low energy field theory. The common assumption is that this effective
field theory at large distance scales must be the Einstein theory. This is a
reasonable supposition if one considers gravitational actions composed of {\it
local} curvature invariants. Then, the Einstein-Hilbert action is the unique
invariant containing no more than two derivatives of the spacetime metric and
would be expected to dominate low energy gravitational physics. In recent works
we have advanced the proposition that on the contrary, at cosmological distance
scales quantum fluctuations of the metric can be important and modify
drastically the classical metric description of general relativity [\cAntMot].

This radical proposal stems from observations about the quantum trace anomaly
of
conformally coupled matter fields in an arbitrary curved background, and the
effective {\it non-local} action it generates for gravitational interactions.
Although the trace anomaly itself is a sum of local terms which are fourth
order
in curvature invariants, the anomalous trace can be described by a coordinate
invariant effective action which is necessarily non-local in the full metric.
Because of this non-locality the trace anomaly has consequences for long
distance physics which may be quite different from that expected in the
classical Einstein theory, and in particular for cosmological spacetimes.
Whatever else a full quantum theory of gravity at the Planck scale entails, the
trace anomaly  due to massless fields remains, and its effects at large
distance
scales must be considered.

The trace anomaly originates from a conflict between coordinate invariance
and scale invariance at the quantum level. This conflict is interesting because
after all, a global dilation or scale transformation is just a particular
coordinate transformation of conformally flat backgrounds (such as the
Friedman-Robertson-Walker spacetime of classical cosmology).  Why should
quantum
physics violate this particular coordinate invariance but no other? Does this
partial breakdown of coordinate invariance survive in the full quantum theory?
How is this global scale symmetry violation related to the infrared behavior of
gravity at very large distance scales?

These questions can be addressed in the effective theory of the conformal
factor
which we have introduced and studied in several recent papers [\cAntMot,\cAMM].
This paper extends and deepens that study by a detailed analysis of
its canonical quantization and physical state space.

In order to isolate the effects of the trace anomaly in the conformal sector of
gravity and study its infrared behavior,  we introduce the conformal
parameterization (or gauge),
$$
g_{ab}(x) = e^{2 \s(x)} \bar g_{ab}(x), \equn\put\confdef
$$
in terms of which the $\s$-dependence of the anomaly induced action becomes
local, as in two dimensions [\cPoly]. When this local $\s$ action is added to
the classical action the total trace anomaly vanishes and scale invariance is
restored in the quantum theory [\cAMM].

In the classical Einstein theory the trace part of the metric is constrained in
terms of the matter sources and cannot fluctuate.  In the effective action for
$\s$ compelled by the trace anomaly this is {\it not} the case, and this new
degree of freedom in the scalar sector is the source of a radical departure
from the classical theory. The fluctuations of $\s$ induced by the trace
anomaly
become relevant  in curved spacetimes at distance scales of the order of the
curvature or greater. This may be seen, for example, by computing the graviton
propagator in de Sitter spacetime, whose logarithmic growth at large spacelike
separations leads to infrared divergences [\cdeS].  These infrared divergences
in gravitational perturbation  theory around cosmological background spacetimes
are the signal of a departure from the classical Einstein theory, and the need
for renormalization  of gravity in the far infrared. Since the conformal
anomaly
records the effects of all massless fields (including gravitons) we have argued
that the renormalization and infrared fixed point of gravity is controlled by
the effective action for $\s$ (``infrared conformal dominance"). The effective
$\s$ theory exhibits some remarkable properties. It is  ultraviolet
renormalizable and possesses a non-trivial infrared stable fixed  point. This
fixed point describes a scale invariant phase of quantum gravity  which is
characterized by certain anomalous scaling relations [\cAntMot].

It is worthwhile to bear in mind that in $D=2$ dimensional gravity the
situation
is similar. There are no degrees of freedom in the Einstein action, which is a
topological invariant in $D=2$.  However, the trace anomaly of matter fields
induces the Polyakov-Liouville action for $\s$ which causes the metric to
fluctuate [\cPoly]. This one new scalar degree of freedom is reflected in the
value of the central charge which is shifted by one unit (from $c_m -26$ to
$c_m
-25$). The fluctuations of $\s$ modify the theory in a radical way through
gravitational dressing of matter fields for example, even at very large
distances where one may have expected quantum gravitational effects to be
small.
Anomalous dimensions and scaling relations  for primary fields and their
correlation functions may be computed, and exhibit behavior qualitatively
different from the classical theory in a fixed metric background [\cKPZ]. In
four dimensions there are in addition transverse  excitations of the metric
which make their own contribution to the trace anomaly. However, even in the
presence of gravitons the effective action for $\s$ has a certain universal
form, dictated by general covariance and the structure of the trace anomaly for
massless fields. Whatever the dynamics of the transverse graviton excitations
the importance of this effective action for $\s$ is that it contains new
dynamics over and above the classical theory which cannot be ignored.

An important difference between $D=2$ and $D=4$ is that the effective action
induced by the trace anomaly of matter fields is quartic in derivatives, and
immediately raises the question of unitarity, which typically plagues theories
with actions with more than two derivatives. The fourth order action contains
both a positive and negative norm scalar at the perturbative level, and the
negative norm scalar can lead to a breakdown of unitarity in the presence of
interactions. However, $\s$ is not just a scalar field but a particular
component of the metric {\confdef}. In a metric theory there are diffeomorphism
constraints which eliminate some of the perturbative excitations. Can one still
make sense of the fourth order $\s$ action as a quantum theory? What are its
physical excitations? Answering these questions is the principal motivation of
this paper. The answer to the first question turns out to be in the
affirmative,
and the determination of the physical states of the $\s$ theory which survive
all the quantum diffeomorphism constraints and their physical interpretation is
our main result.

Before entering into the physical state conditions in a fourth order theory it
is worthwhile to recall that even the second order Einstein-Hilbert Lagrangian
exhibits a negative norm scalar which leads to an unbounded Euclidean action if
treated in a naive fashion. In the classical Einstein theory this field is
non-propagating (except possibly for a finite number of modes in spacetimes
with
certain topologies). It can be eliminated from the theory by the constraints of
coordinate invariance [\cYork], which in a canonical framework are just the
$T^{0i}$ and $T^{00}$ components of the equations of motion. In the Einstein
theory this may be understood by simple counting of degrees of freedom. One
begins with a real, symmetric metric which contains ten degrees of freedom.
Four
of these are pure gauge, corresponding to the general four vector
parameterizing
infinitesimal diffeomorphisms. In a canonical framework the vanishing of the
momenta corresponding to these four invariances lead to four secondary first
class constraints, which reduce the number of true propagating degrees of
freedom to $10-4-4=2$ transverse gravitons in the Einstein theory. The negative
norm scalar $\s$ of the Einstein theory has dropped out, or more precisely, it
is constrained and determined by the matter sources. When the trace anomaly is
taken into account this counting of degrees of freedom is modified by the
appearance of a quartic action for $\s$, which (since the classical action was
already second order) has the effect of introducing one additional degree of
freedom into the quantum theory in an analogous way to the Liouville theory in
two dimensions, which shifts $c_m -26 \rarr c_m -25$. The question now becomes:
what physical states does this additional degree of freedom contain over and
above the gravitons of the low-energy Einstein theory?

In this paper we study the canonical quantization and the issue of unitarity
of the quartic effective theory induced by the trace anomaly in four
dimensions.
We quantize the effective action of the $\s$ field in terms of canonical
commutation relations of positive and negative metric in the Fock space of
modes propagating in the spacetime background $R \times S^3$. Then,
following the methods of the previous paper [\cpapI] (hereafter denoted as
Paper
I in this work) we compute the moments of the energy-momentum tensor deduced
from the same effective $\s$ action and derive the constraints of spatial
diffeomorphisms on the three sphere. We then apply the positive frequency part
of these constraints to the Hilbert space of states in the theory, level by
level, and determine the physically  allowed states in the Fock space which
satisfy all the constraints of spatial diffeomorphism invariance. Lastly, we
apply the quantum Hamiltonian constraint with the finite shift determined in
Paper I. Although there is {\it no} propagating local degree of freedom
(negative metric or otherwise) we find an infinite tower of {\it discrete}
states which satisfy all the constraints. It should be emphasized that all
these
invariant states in the pure $\s$ theory are present independently of the
ordinary local spin two excitations of the metric.

Depending on the value of the anomaly coefficient we can distinguish two cases.
In the first case the Hamiltonian constraint is satisfied with a real momentum
eigenvalue of the zero mode of $\s$ in this background. In this case,
corresponding to $c_m > 25$ in $D=2$, we can carry out the analysis of
unitarity
and show that the discrete physical states have manifestly positive norm.
In the second case the Hamiltonian constraint is satisfied with an imaginary
momentum eigenvalue,  corresponding to $c_m < 1$ in two dimensions. In this
case
the physical states are non-normalizable but instead correspond to operators of
scalar observables. In fact, in the semiclassical limit, they are created by
integer powers of the Ricci scalar operator integrated over spacetime. Because
of the trace anomaly, the Ricci scalar is {\it not} constrained to be a
constant
by the equations of motion in the quantum theory and powers of this quantity
correspond explicitly to different allowed physical states.

The outline of the paper is as follows. In Section 2 we introduce the canonical
quantization of the $\s$ field in terms of Fock space oscillators with both
positive and negative metric on $R \times S^3$. The advantages of quantizing in
the background Einstein universe and the general framework are laid out in
detail in paper I, the Appendix of which contains many relations involving the
spherical harmonic functions which we make extensive use of in this paper as
well. From the energy-momentum tensor for the $\s$ field we derive detailed
expressions for the moments of its $T^{0i}$ components with the  volume
non-preserving and volume preserving harmonics on $S^3$. In Section 3 we apply
the positive frequency part of these moments to the possible states level by
level in the Fock space. A particular linear combination of Fock space
operators
raised to the $n^{th}$ power yields a physical state at each even level $2n$,
which survives the application of all the constraints. In Section 4, the
correspondence of the tower of allowed physical states with operators of
well-defined scaling dimension in four dimensions is presented. We also repeat
the analysis of physical states with the addition of conformally coupled scalar
matter, and show that the interpretation is robust under the addition of matter
fields. Finally, Section 5 contains our main conclusions and a discussion of
the
results.

\vfill
\eject

\newcount\eqncount
\sectadv
\set\sect{2}
\beginsection{2. Quantization of $\s$}

In paper I we have presented the general framework of canonical quantization on
$R\times S^3$, the diffeomorphism constraints and the correct subtraction
of the Hamiltonian $H$ determined by the ghost system. This discussion applies
to any theory with general coordinate invariance and zero stress-tensor trace.
We examine now the particular case of the theory of the conformal factor
induced
by the trace anomaly. The induced effective action for $\s$ in spacetimes of
Lorentzian signature is [\cAntMot]
$$
\G = -{Q^2 \over (4 \pi)^2}\int d^4 x \sqrt {-g} \bigl[
\s  \D_4 \s + \hbox{$1\over 2$}\bigl(G - \hbox{$2 \over 3$} \sq R \bigr) \s
\bigr]\ ,
\equn\put\Seff
$$
where the Weyl covariant fourth order operator is
$$\eqalign{
\D_4 &\equiv \sq ^2 + 2R^{ab} \de_a \de_b - \hbox{$2 \over 3$} R {\sq}+
\hbox{$1 \over 3$}(\de^a R)\de_a \cr
&= \sq^2 + 4\,\pa_t^2 \quad {\rm on } \quad R \times S^3 \ ,\cr}
\equn\put\fourth
$$
with unit radius, so that $R_{0a}=0, R_{ij} = 2 \d_{ij}, R= 6$.
The parameter $Q^2$ is proportional to the coefficient of the
Gauss-Bonnet term $G = \Rie - 4 \Ric + \R $ in the general form of the trace
anomaly for conformal fields, and therefore depends on the matter content of
the
theory coupled to $\s$. Although it is known that $Q^2>0$ for all free matter
fields we treat $Q^2$ as a free parameter in the following development.

The equation of motion for $\s$ derived from the action $\G$ is
$$\eqalign{
\D_4 \s &= -\hbox{$1\over 4$}\bigl(G - \hbox{$2 \over 3$} \sq R \bigr)\cr
&= 0 \quad {\rm on } \quad R \times S^3 \ ,\cr}
\equn\put\eoms
$$
because $G = \sqR = 0$ on $R \times S^3$. Hence, $\s$ is a free field which may
be expanded in functions of the form, $\exp (-i\o t) \Y$ with $\Y$ the
scalar harmonics of the 3-sphere, for which the operator $\D_4$ factorizes as
$$
\D_4\left\{\exp (-i \o t) \Y\right\} = \left(\o^2-(2J)^2)\right)\left(\o^2
- (2J+2)^2\right) \exp (-i \o t) \Y \ .\equn
$$
The sourcefree wave equation and simple factorization of the fourth order
operator $\D_4$ are important simplifications that hold on the Einstein
universe
$R\times S^3$. The two factors correspond exactly to two independent free
fields
which we shall designate $\s_1$ and $\s_2$, respectively, each obeying second
order wave equations. Each of these fields may be expanded in terms of
canonical
creation and destruction operators:
$$\eqalign{
\s_1 =& {\p \over Q} \mathop{{\sum_{JM}}'} {1\over\sqrt{2J(2J+1)}}
\left(e^{-2iJt} \Y a_{_{JM}} + e^{2iJt} \Ys a^{\dag}_{_{JM}}\right)\cr
\s_2 =& {\p \over Q} \sum_{JM} {1\over
\sqrt{2(2J+1)(J+1)}}\left(e^{-2i(J+1)t}\Y
b_{_{JM}} + e^{2i(J+1) t} \Ys b^{\dag}_{_{JM}}\right)\ ,\cr}
\equn\put\smod
$$
where the prime denotes the omission of the $J=0$ mode from the first
oscillator
sum. Its place is taken by the zero mode with linear time dependence, {\it
i.e.}
$$
\s = \s_0 + \s_1 + \s_2 \ ,\quad {\rm with}\quad \s_0 = {1\over Q}
(\hat q + \hat pt)\ .
\equn\put\allm
$$
The normalization of the modes in {\smod} is fixed by the normalization of the
action {\Seff}. If the oscillator creation and destruction operators are
normalized in the usual way,
$$\eqalign{
\left[a_{_{J_1M_1}}, a^{\dag}_{_{J_2M_2}}\right] &=
\d_{_{J_1J_2}}\d_{_{M_1M_2}}\cr
\left[b_{_{J_1M_1}}, b^{\dag}_{_{J_2M_2}}\right] &=
-\d_{_{J_1J_2}}\d_{_{M_1M_2}} \qquad {\rm for}\quad Q^2 > 0\ ,\cr}
\equn\put\comm
$$
then we find that the Feynman function of the full $\s$ field satisfies
the correctly normalized inhomogeneous wave equation,
$$
i\D_4 \langle T\s(x)\s(x')\rangle =  {8\p^2\over Q^2}\d^4(x,x')\ ,
$$
provided that the zero mode coordinate and momentum obey
$$
[\hat p, \hat q] = -i\ .\put\pqc
\equn\put\pqc
$$

Because the action $\G$ is fourth order in derivatives, there are necessarily
both positive and negative norm states in the Fock space generated by the
$a^{\dag}$ and $b^{\dag}$, and the classical energy-momentum of the theory is
not positive.  The minus sign in the second of the commutation relations
{\comm}
signifies that $\s_2$ is quantized with {\it negative} metric if $Q^2 > 0$
while
$\s_1$ is a positive metric field. Conversely, if $Q^2 < 0$ then $\s_1$ should
be quantized with negative metric, while $\s_2$ becomes the positive metric
field, and the signs in {\comm} and {\pqc} should be reversed. Hence, in either
case there are certainly negative norm states in the Fock space prior to the
imposition of the diffeomorphism constraints. In order to find these
constraints
we need the energy-momentum tensor corresponding to the action $\G$.
This normal ordered energy-momentum tensor is given by
$$\eqalign{
-{(4\p)^2 \over Q^2}T^{ab}&= :2\sq (\de^{a}\s \de^{b}\s) + \hbox {$4\over 3$}
\de^{a}\de^{b}\left(\de^{c} \s\de_{c}\s\right) + 4\de^{(a}\s\de^{b)}\sq\s
-4\de_c\de^{(a}(\de^{b)}\s\de^c\s)\cr
& -\hbox {$2\over 3$}\de^a\de^b \sq \s + 8 R^{(a}_c\de^{b)}\s \de^c\s
- \hbox {$4\over 3$} R^{ab}\left(\de^{c} \s\de_{c}\s\right)
- \hbox {$4\over 3$} R \de^a\s\de^b\s\cr
& + \hbox {$2\over 3$}R^{ab} \sq \s + \hbox {$2\over 3$}\s \de^a\de^b R
+ 2\de^a\de^b(R\s) + 4 \sq (R^{ab}\s) -8\de^c\de^{(a}(R^{b)}_c\s)\cr
& + 4\de_c\de_d\left( R^{c(ab)d}\s \right) -\hbox {$2 \over 3$}
\de^{(a}\left( \s \de^{b)}R\right) -2 R^a_{cde}R^{bcde}\s + 8 R^{ac}R^b_c\s -
2R^{ab} R\s \cr
&+ g^{ab} \Bigl\{ (\sq\s)^2 -\hbox {$1\over 3$} \sq
\left(\de^{c} \s\de_{c}\s\right) - 2R^{cd}\de_c\s\de_d\s + \hbox
{$2\over 3$}R \left(\de^{c} \s\de_{c}\s\right) + \hbox {$2\over 3$}\sq^2\s \cr
& - 2\sq (R\s) + 4\de^c\de^d (R_{cd}\s) + \hbox {$1\over 3$}\de^c(\s\de_cR) +
\hbox {$1\over 2$} (G - \hbox {$2\over 3$}\sq R)\s\Bigr\}:\ .\cr}
\equn\put\Tab
$$
Its trace is simply proportional to the $\s$ equation of motion,
$$
T^a_a = -2 {Q^2\over (4\p)^2 } \D_4\s = {Q^2\over 32\p^2}\bigl(G - \hbox{$2
\over
3$} \sq R \bigr) = 0 \ ,\equn\put\strace
$$
which vanishes by {\eoms}. This is no accident, of course, but is
the result of the total anomaly cancellation in the $\s$ theory.
Just as in two dimensions the ``classical" trace from the linear term in the
action $\G$ is opposite in sign from the trace anomaly of matter fields which
gave rise to the effective $\s$ action {\Seff} in the first place. Hence
the trace of the $\s$ energy-momentum tensor is precisely equal to minus
the $\s$ variation of the action {\Seff} which vanishes on $R \times S^3$.

On $R\times S^3$ the energy density may be written in the form,
$$\eqalign{
-{(4\p)^2 \over Q^2}T^{00}[\s]\big\vert_{_{R\times S^3}} &= :2\dot\s\sq\dot\s
-2 \ddot\s\sq\s - \ddot\s\tri\s -(\sq\s)^2 +\s\tri\ddot\s -4\dot\s^2\cr
&\quad +\tri\left(\hbox {$2\over 3$} \dot\s^2 -\s\ddot\s + \hbox {$1\over
6$}\tri \s^2 - \hbox {$1\over 3$} \s\tri \s - \hbox {$2\over 3$} \sq\s +
4\s\right):\ ,
\cr}\equn\put\enden
$$
where $\tri \equiv \nabla^i\nabla_i$ is the Laplacian operator on $S^3$.
The generator of the spatial diffeomorphism constraints is
$$\eqalign{
-{(4\p)^2 \over Q^2} T^{0i}[\s]\big\vert_{_{R\times S^3}} &=
 :-2\dot\s \de^i\sq\s - 2(\sq\dot\s) (\de^i\s) + 2 \de^i\de^c (\dot\s\de_c\s)
+ 2 \pa_t\de^c (\de^i\s \de_c\s)\cr
-& 2\sq(\dot\s\de_i\s) - \hbox {$4 \over 3$}\pa_t\de^i (\de^c\s\de_c\s) -
4\de^i\dot\s + \hbox {$2 \over 3$} \de^i\sq\dot\s + 4\dot\s\de^i\s:\ .\cr}
\equn\put\spa
$$

Substituting the mode expansion {\smod} into {\enden} and forming the moments
of
the volume non-preserving diffeomorphisms with positive frequency part,
$$
\dT \equiv -i\sqrt{V}\int_{S^3}\, d\O (\de_i \Ys )\, T^{0i (+)} =
\sqrt{V} \int_{S^3}\, d\O (\de_i \Ys )\, \pa_t T^{00 (+)}
$$
yields the following somewhat formidable expression for the moments:
$$\eqalign{
&\dT\big\vert_{_{t=0}} =  -\hbox {$2J\over 3$}
\sqrt{\hbox {$J\over 2J+1 $}}(J+1)(2J+3)(Q-i\hat p)\  a_{_{JM}}\cr
&\qquad\qquad + \hbox {$2J\over 3$}\sqrt{\hbox {$J+1\over 2J+1$}}(J+1)(2J
-1)(Q+i\hat p)\  b_{_{JM}}\quad + \cr
&\sum_{\jo\mo}\sum_{\jt\mt}\cc^{JM}_{\jo\jt\mo\mt}\Biggl\{\hbox
{$(\jo +\jt)\over 2\sqrt {\jo\jt}$} a_{_{\jo\mo}}a_{_{\jt\mt}}
\Bigl\{ (\jo -\jt)^2(\jo + \jt) (\jo + \jt +1)\cr
&\quad + {\hbox{$J(J+1)\over 3$}}\left[ 4\jo\jt - 4(\jo^2 +
\jt^2) - (\jo + \jt) + J(J+1)\right]\Bigl\}_{_{\jo > 0;\jt >0}}\cr
&+ \hbox {$(\jo +\jt +1)\over\sqrt{\jo(\jt + 1)}$}\ a_{_{\jo\mo}}b_{_{\jt\mt}}
\Bigl\{ 4(\jo-\jt)\jo(\jt+1) + (\jo -\jt)(\jo + \jt +1)^2 (\jo-\jt - 1)\cr
&\quad + {\hbox{$J(J+1)\over 3$}}
\left[ 4(\jo\jt - \jo^2 - \jt^2) + 3\jo - 7\jt - 3 + J(J+1) \right]
\Bigr\}_{_{\jo > 0}}\cr
&+ \hbox {$(\jo + \jt + 2)\over 2\sqrt{(\jo+1)(\jt+1)}$}\
b_{_{\jo\mo}}b_{_{\jt\mt}}\Bigl\{(\jo-\jt)^2(\jo+\jt+1) (\jo+\jt +2) +\cr
&\quad \hbox{$J(J+1)\over 3$} \left[4(\jo\jt - \jo^2 - \jt^2) - 3(\jo + \jt) -
2
+ J(J+1) \right]\Bigr\}\cr  }
$$
$$\eqalign{
&+ \hbox {$(\jo-\jt) \over \sqrt{\jo\jt}$}\ {\tilde
a}^{\dag}_{_{\jt\mt}} a_{_{\jo\mo}}\Bigl\{-4\jo\jt(\jo + \jt +1) +
(\jo-\jt)^2(\jo +\jt)(\jo +\jt +1)\cr  &\quad - {\hbox{$J(J+1)\over 3$}}
\left[4(\jo\jt + \jo^2 +\jt^2) + \jo +\jt -J(J+1)\right]
\Bigr\}_{_{\jo >\jt >0}}\cr
& +\hbox {$(\jo-\jt -1) \over \sqrt{\jo (\jt +1)}$}\
{\tilde b}^{\dag}_{_{\jt\mt}} a_{_{\jo\mo}}
\Bigl\{(\jo -\jt)(\jo - \jt -1)(\jo +\jt +1)^2  - \cr
&\quad {\hbox{$J(J+1)\over 3$}}\left[ 4(\jo\jt + \jo^2 +\jt^2) + 5\jo + 7\jt +
3
- J(J+1) \right]\Bigr\}_{_{\jo > \jt + 1}}\cr
&+\hbox {$(\jo-\jt +1) \over
\sqrt{(\jo + 1)\jt}$}\   {\tilde a}^{\dag}_{_{\jt\mt}} b_{_{\jo\mo}}
\Bigl\{(\jo -\jt)(\jo - \jt +1)(\jo +\jt +1)^2 -\cr
&\quad {\hbox{$J(J+1)\over 3$}}\left[4(\jo\jt + \jo^2 +\jt^2) + 7\jo + 5\jt
+ 3 -
J(J+1) \right]\Bigr\}_{_{\jo + 1 >\jt > 0}}\cr &+ \hbox{$(\jo-\jt) \over
\sqrt{(\jo + 1)(\jt +1)}$}\  {\tilde b}^{\dag}_{_{\jt\mt}}  b_{_{\jo\mo}}
\Bigl\{(\jo +\jt + 1)\left[4(\jo + 1)(\jt +1) +
(\jo-\jt)^2(\jo +\jt +2)\right]\cr
&\quad  - \hbox {$J(J+1)\over 3$}
\bigl[4(\jo\jt + \jo^2 +\jt^2) + 11(\jo + \jt) + 10 - J(J+1)
\bigr]\Bigr\}_{_{\jo >\jt}}\Biggr\}\ .\cr }
\equn\put\mons
$$
Here the notation [\cpapI]
%
%
$$
\eqalign{
{\tilde a}_{JM} &\equiv \e_{_M} a_{J\, -M}\cr
{\tilde b}_{JM} &\equiv \e_{_M} b_{J\, -M} \qquad {\rm with}\cr
\qquad \e_{_M}&\equiv (-)^{m-m'}\cr}
\equn
$$
has been used, and the symbol $\cc$ is an $O(4)$ angular momentum coupling
coefficient that may be expressed as a product of two ordinary $SU(2)$
Clebsch-Gordon coefficients $C^{Jm}_{\jo m_1\jt m_2}$ via
$$\eqalign{
\cc^{JM}_{\jo\mo\jt\mt} &\equiv  {\sqrt {V}\over \sqrt{(2\jo+1)(2\jt + 1)}}
\int_{S^3} d\O\  \Ys Y_{_{\jo\mo}} Y_{_{\jt\mt}}\cr &= {1\over \sqrt{(2J + 1)}}
C^{Jm}_{\jo m_1\jt m_2}C^{Jm'}_{\jo m'_1\jt m'_2}\ .\cr}
\equn
$$

The moments of the volume preserving diffeomorphisms may be constructed
by substituting the mode expansions {\smod} into {\spa} and computing
$$
X^{(+)}_{_{J\cm}} \equiv -\int_{S^3}\,d\O\, \cy_{i_{\,J\cm}}\, T^{0i(+)}\
,\equn
$$
where the $\cy^i_{J\cm}$ are the transverse vector harmonics on $S^3$ as
defined
in paper I. In a calculation analogous to that leading to {\mons} we find
$$\eqalign{
&X^{(+)}_{_{J\cm}}\big\vert_{_{t=0}} =  {-i\over 4}
\mathop{{\sum_{\jo\mo}}'}\mathop{{\sum}'}_{\jt\mt}
\ck^{JM}_{\jo\mo\jt\mt}\biggl\{\hbox {$(\jo -\jt)\over 2\sqrt {\jo\jt}$}
a_{_{\jo\mo}}a_{_{\jt\mt}} \left[ 4J(J+1) - 4(\jo +\jt)^2 + 1 \right]\cr  &+
\hbox {$1\over\sqrt{\jo(\jt + 1)}$}\ a_{_{\jo\mo}}b_{_{\jt\mt}}
\bigl\{ \left[4J(J+1) - 4(\jo + \jt + 1)^2 + 1 \right](\jo - \jt - 1) -
16\jo(\jt
+ 1)\bigr\}\cr &+ \hbox {$(\jo - \jt)\over 2\sqrt{(\jo+1)(\jt+1)}$}\
b_{_{\jo\mo}}b_{_{\jt\mt}}
\left[4J(J+1) - 4(\jo + \jt + 2)^2 +1 \right]\Bigr\}\cr
&+ \hbox {$1\over \sqrt{\jo\jt}$}\
{\tilde a}^{\dag}_{_{\jt\mt}} a_{_{\jo\mo}}
\bigl\{\left[4J(J+1) - 4(\jo - \jt)^2 + 1\right](\jo +\jt) + 16\jo\jt
\bigr\}_{_{\jo \ge\jt}}\cr
&+ \hbox {$(\jo + \jt + 1) \over \sqrt{\jo (\jt +1)}$}\
{\tilde b}^{\dag}_{_{\jt\mt}} a_{_{\jo\mo}}
\left[4J(J+1) - 4(\jo - \jt - 1)^2 + 1 \right]_{_{\jo \ge \jt + 1}}\cr
&+\hbox {$(\jo + \jt +1) \over \sqrt{(\jo + 1)\jt}$}\
{\tilde a}^{\dag}_{_{\jt\mt}} b_{_{\jo\mo}}
\left[4J(J+1) - 4(\jo - \jt + 1)^2 + 1\right]_{_{\jo + 1 \ge \jt}}\cr
&+ \hbox {$1 \over \sqrt{(\jo + 1)(\jt +1)}$}\
{\tilde b}^{\dag}_{_{\jt\mt}}  b_{_{\jo\mo}}
\bigl\{\bigl[4J(J+1) - 4(\jo - \jt)^2 + 1\bigr] (\jo + \jt + 2)\cr
& \qquad - 16(\jo + 1)(\jt + 1)\bigr\}_{_{\jo \ge\jt}}\biggr\}\ ,\cr}
\equn\put\monspa
$$
where the coupling coefficient with the transverse vector $O(4)$ spherical
harmonic $\cy^i_{_{J\cm}}$ is defined by
$$
\ck^{J\cm}_{\jo\mo\jt\mt} \equiv {1\over 2\sqrt{(2\jo+1)(2\jt + 1)}}
\int_{S^3} d\O\  \cy^i_{_{J\cm}}\, Y_{_{\jo\mo}}\nabla_i Y_{_{\jt\mt}}\ .
\equn\put\vcoup
$$

The lowest moments of the diffeomorphism generators are just the $SO(4,2)$
conformal group generators which now have an explicit realization in terms of
the $\s$ field creation and destruction operators. The lowest non-trivial
moment
of the volume non-preserving diffeomorphism generators is
$$
\eqalign{
K_{_M}^{(+)} &= \dot T_{\ha M}^{(+)}\Big\vert_{_{t=0}}=
-(Q-i\hat p)\, a_{\ha M} +\cr
& \sqrt{2} \sum_{JM_1M_2} \cc^{\ha M}_{J+\ha M_1 J M_2}\Bigl\{
-2(J+1)\sqrt{J(2J+1)} {\tilde a}^{\dag}_{_{J - M_2}} a_{_{(J+\ha)\, M_1}}\cr
&+ (2J+1)\sqrt{(J+1)(2J+3)}\,{\tilde b}^{\dag}_{_{J - M_2}} a_{_{(J+\ha)\,
M_1}} -\sqrt{(J+1)(2J+1)}\,{\tilde a}^{\dag}_{_{(J +\ha)\, -M_1}} b_{_{J M_2}}
\Bigr\}\ ,\cr }
\equn\put\momha
$$
which together with its Hermitian conjugate form the $8$ special conformal
generators of $R\times S^3$.

The lowest moments of the volume preserving diffeomorphisms are precisely
the $6$ rotation generators of $SO(4)$,
$$
R_{_{M_1M_2}} = -i \sum_{JM_3M_4} \int d\O\, \r^i_{_{M_1M_2}}
\left(Y_{_{JM_3}}\nabla_iY_{_{JM_4}}^*\right) \left(a^{\dag}_{_{J M_4}}
a_{_{J M_3}} -\, b^{\dag}_{_{J M_4}} b_{_{JM_3}}\right)\ . \equn
$$

With these expressions for the $8 +6 = 14$ generators of the global conformal
group in hand, it is straightforward to evaluate their commutators, and compare
to the form expected from the classical Lie algebra of $SO(4,2)$. In
particular,
the quantum commutator
$$
\left[ K^{(+)}_{_{M_1}}\,, K^{(-)}_{_{M_2}}\right]
= 2\d_{_{M_1M_2}}H  + 2R_{_{M_1M_2}}\ ,
\equn\put\shifts
$$
yields the Hamiltonian operator $H$,
$$
H = \ha \hat p^2 + \ha Q^2 + 2\sum_{JM}\left\{ Ja^{\dag}_{_{JM}}a_{_{JM}} -
(J+1)b^{\dag}_{_{JM}}b_{_{JM}}\right\}
\equn\put\Ham
$$
which differs from the spatial integral of the normal ordered $:T^{00}:$
by a well-defined c-number shift, $Q^2/2$.

The additional $\hbox {$1 \over 2$}Q^2$ contribution is the analog of that
obtained in two dimensional quantum gravity [{\cPoly},{\cLiou}]. In $D=2$ the
Polyakov-Liouville effective action has the fully covariant non-local form
[\cPoly],
$$
{Q^2\over 16 \pi}\int d^2x \sqrt {-g}\int d^2x'\sqrt{-g'} \ R(x)
\sq ^{-1}(x,x') \ R(x')\ ,
\equn\put\loctwo
$$
where
$$
Q^2 = {25-c_m\over 6}\ , \qquad D=2\equn\put\qdef
$$
is fixed in terms of the matter central charge $c_m$, by the anomaly
cancellation condition at the conformal fixed point. By calculating the stress
tensor of the invariant action {\loctwo} we find an additional contribution to
the energy density of the $\s$ action which on $R\times S^1$ results in a
constant c-number shift in the Hamiltonian equal to $Q^2/2$ on the cylinder.
Alternatively, the shift is obtained just as readily from the $D=2$ global
conformal algebra $SO(2,2) \cong SL(2, C)$ by computing the commutator $[L_1,
L_{-1}]$ in the analog of {\shifts}.

In $D=4$ the fully covariant but non-local action is [\cAntMot, \cAMM],
$$
{Q^2 \over 4\p^2} \int d^4 x \sqrt {-g}  \int d^4 x' \sqrt {-g'} (G -
\hbox{$2\over 3$}\sq R)(x)  (\D_4)^{-1}(x,x') (G - \hbox{$2 \over 3$} \sq
R)(x') .\equn\put\nonf
$$
When the conformal parameterization {\confdef} is substituted into this
non-local action we obtain {\Seff} plus an additional $\s$ independent term
of exactly the same form as {\nonf} but evaluated in the background metric
$\bar
g_{ab}$. Varying this additional $\s$ independent action with respect to the
background metric gives an additional background dependent term in the
energy-momentum tensor of the form,
$$\eqalign{
{\overline T}^{ab} &= \hbox {$\a \over (4\p)^2$}\left[ g^{ab}( \hbox {$1
\over 2$} R^2 - R_{cd}R^{cd} ) + 2 R^{ac}R^b_c - \hbox {$4 \over 3$} R R^{ab}
\right]\cr & + \hbox {$\b \over (4\p)^2$}\left[ -\hbox {$1 \over 18$}g^{ab}
(R^2 - 4\sq R) + \hbox {$2 \over 9$} R R^{ab} - \de^a\de^b R \right]\ ,\cr}
\equn\put\clT
$$
with $\a=-\b=-\hbox {$1 \over 2$}Q^2$. On $R\times S^3$
with unit radius we obtain, in particular, the additional contribution
$$
{\overline T}^{00} \big \vert_{_{R\times S^3}} = {Q^2 \over 4\p^2}\ .
\equn\put\clen
$$
Integrating this result over the unit $S^3$ with volume $V = 2\p^2$ gives
precisely the $\hbox {$1 \over 2$}Q^2$ contribution to the total Hamiltonian in
{\Ham}. Because $R \times S^3$ is a product space there is no shift
contribution to $T^{0i}$ in eq. {\spa}.

It is also interesting to remark that the c-number shift in the Hamiltonian
generator on the cylinder results from anomaly cancellation in the full quantum
theory of matter plus $\s$ plus ghosts in $D=2$ and that this relation has a
four dimensional analog as well. In fact the c-number shift is just the vacuum
Casimir energy of all these fields  on the manifold $R\times S^1$.
The Casimir energy is determined by the trace anomaly coefficient
for all fields [\cLBC], {\it except} the ghosts, and one finds
$$
\langle 0\vert T^{(qu)\,00}\vert 0\rangle \big \vert_{_{R\times S^1}}
= -{1\over 24\p} \left(c_m + 1 + (-26+24)\right) \equn\put\ttwo
$$
Now using the anomaly cancellation condition {\qdef} and integrating over $S^1$
one finds for the total zero-point energy, $Q^2/2 - 2$, where the shift $-2$
arises from the additional contribution of ghosts $24$ in {\ttwo}. That the
ghost energy has a contribution in addition to its trace anomaly coefficient
is a reflection of the fact that the ghost action arises from gauge-fixing and
is therefore not fully coordinate invariant. The additional $-2$ shift can be
computed for instance by evaluating the commutator $[L^{gh}_1, L^{gh}_{-1}]$ in
the ghost sector [\cGSW].

In four dimensions the trace anomaly has the general form
$$
\langle 0\vert T^{(qu)\,a}_a\vert 0\rangle  = b C^2 +  b'G +
(\hbox {$2\over 3$}b+b'') \sq R\ ,
\equn
$$
where $C^2$ is the Weyl tensor squared. As in $D=2$, for conformally flat
backgrounds (for which $C^2=0$), the expectation value of the full
energy-momentum tensor of conformally invariant matter may be determined from
the trace anomaly [\cLBC]. In $D=4$ the result is identical to {\clT} with $\a
=
16\p^2 b'$ and $\b = 8\p^2 (2b + 3b'')$. Integrating this result over the unit
$S^3$ we obtain for the additional c-number shift in the vacuum energy,
$$
\int_{S^3} d\O\,\langle 0\vert T^{(qu)\, 00}\vert 0\rangle \big
\vert_{_{R\times S^3}} = 2\p^2(-6b' + 2b + 3b'')\
\equn\put\vaceng
$$
Now, the conformal fixed point condition in four dimensions enforces a
constraint
on the anomaly coefficients [\cAntMot,\cAMM]
$$
2b' + 2b + 3b'' = 0
\equn
$$
for the total of all fields including $\s$. This relation is a consequence
of the condition that the $\beta$-function of the $R^2$ coupling in the
effective action vanishes at the infrared fixed point. In addition, anomaly
cancellation at the fixed point requires
$$
Q^2 = -32\p^2 b'\ , \qquad D=4
\equn
$$
which is the analog of {\qdef} in $D=2$. Hence we find for the total Casimir
energy
$$
\int_{S^3} d\O\, \langle 0\vert T^{(qu)\, 00}\vert 0\rangle \big
\vert_{_{R\times S^3}}
= \hbox {$1\over 2$}Q^2 - 4
\equn
$$
where the $-4$ is the additional contribution of the ghost computed in
paper I and the analog of $-2$ in two dimensions.

The explicit form of the moments of the $\s$ energy-momentum tensor given by
eqs. {\mons} and {\monspa}, together with the quantum Hamiltonian {\Ham} and
its
finite shift are the elements we need to determine the physical states in the
quantum $\s$ theory in the next section.

\vfill
\eject

\newcount\eqncount
\sectadv
\set\sect{3}
\beginsection{3. The Physical States}

The physical states of the pure $\s$ theory will be determined now by
applying the physical state conditions [\cpapI],
$$\eqalign{
\dT \vert phys\rangle &= 0\cr
X^{(+)}_{J\cm}\vert phys\rangle &= 0\cr
H \vert phys\rangle &= 4\vert phys\rangle\cr}
\equn\put\qphys
$$
to all possible states in the Fock space. We note first that by considering
only
the energy-momentum tensor of the pure $\s$ action we shall be determining the
physical spectrum of only the conformal sector of quantum gravity in the vacuum
sector of any matter or transverse, graviton excitations we may wish to add in
a
more complete theory. We have nothing new to add about the inclusion of
gravitons in the conformally invariant quantization scheme of this paper, and
remark only that if included they will generate new physical states in the
transverse sector over and above what we find in the pure $\s$ theory. However
any states we find in the pure $\s$ theory will survive also in the full theory
with the matter and/or transverse graviton sectors.

Before applying the quantum constraints let us make some observations about the
purely classical $\s$ theory. Classically, one may use the mode expansions
{\smod} and {\allm} with all quantities ordinary c-numbers. Because of the
fourth order operator $\D_4$ there are two sets of wave modes parameterized by
$a$ and $b$. There are no exponentially growing solutions because $\o^2\ge 0$
for both sets of modes. However for $Q^2 > 0$ the energy density of all the $b$
modes is negative in {\Ham}. When all the classical diffeomorphism constraints
are imposed $T^{0i} = T^{00}=0$ these negative energy solutions are eliminated
and there is no instability at the classical level (in the free $\s$ theory).

In the quantization scheme of the previous section we have traded negative
energy states for negative norm states. Since the quantum constraints on the
physical states {\qphys} are weaker than the corresponding classical
conditions,
the elimination of unphysical negative norm states by the quantum
diffeomorphism
constraints is a non-trivial test that the quantum $\s$ theory must pass.

Let us begin by observing that the Fock vacuum $\vert0\rangle$ of the $a$
and $b$ oscillators is characterized by an additional quantum number which
is the eigenvalue of the zero mode momentum operator
$$
\hat p \vert 0, p \rangle = p \vert 0, p \rangle \ .
\equn
$$
The general state is constructed by operating with any number of
$a^{\dag}$ and $b^{\dag}$ creation operators on a Fock vacuum of this kind.
{}From the form of the Hamiltonian {\Ham} we observe that each
$a^{\dag}_{JM}$ contributes $2J$ to the energy while each $b^{\dag}_{JM}$
contributes $2J +2$, both of which are integers. Hence, we define the integer
level $N$ of the general $\s$ state by
$$
H \vert N, p \rangle = \left( \ha p^2 + \ha Q^2 + N\right)
\vert N, p \rangle \ .
\equn\put\level
$$
Since terms with different $N$ cannot cancel, we can apply the $\dT$
constraints
to each $N$ level independently. The application of the volume non-preserving
conditions in the pure $\s$ theory now proceeds in a way analogous to that
described in paper I for a scalar matter field, level by level.

The ground state at level $0$ clearly satisfies the volume non-preserving
constraints,
$$
\dT \vert 0, p \rangle = 0\ ,
\equn
$$
and is therefore a candidate physical state.
For $N=1$ the only possible state involves a linear combination of
$$
a^{\dag}_{\ha M}\vert N=1, p\rangle \ .
$$
Applying $\dot T_{\ha M}^{(+)}$ to this is non-vanishing unless all the
coefficients of this linear combination vanish identically. Hence there are
no physical states at level one.

At level two, the general state is of the form,
$$
\vert 2, p\rangle = \left(\a_{_{\mo\mt}} a^{\dag}_{\ha \mo}a^{\dag}_{\ha
\mt} + \b b^{\dag}_{000} + \g _{_M}a^{\dag}_{1M}\right)\vert 0, p\rangle \ .
\equn
$$
Applying  $\dot T_{{3\over 2} M}^{(+)}$ to this state gives a non-vanishing
contribution unless
$$
\g_{M} = 0
\equn
$$
identically.  Applying $\dot T_{1M}^{(+)}$ gives no new constraint because the
coefficient of $a_{_{\ha}}a_{_{\ha}}$ vanishes for $J=1$. Finally the
$\dot T_{\ha M}^{(+)}$ gives the non-trivial condition,
$$
\left[2(Q-ip) \a_{_{M\mo}} a^{\dag}_{\ha
\mo}- \e_{_M}\b a^{\dag}_{\ha\, -M}\right]\vert 0, p\rangle = 0
$$
which is satisfied if
$$\eqalign{
i) & Q=ip \quad {\rm and}\quad \b=0 ,\quad {\rm or}\cr
ii)& Q\ne ip \quad {\rm and}\quad \a_{\mo\mt} = {\b
\over 2 (Q-ip)} \e_{_{\mo}} \d_{\mo \,-\mt}\ .\cr }
\equn\put\conds
$$
We see that there is a non-trivial state surviving the application of the
volume non-preserving spatial diffeomorphism constraints at level $2$.

At level three the general state is created by a linear combination of the five
operators, $a^{\dag}_{{3 \over 2}}$, $b^{\dag}_{\ha}$,
$a^{\dag}_{1}a^{\dag}_{\ha}$, $b^{\dag}_{0}a^{\dag}_{\ha}$, and
$a^{\dag}_{\ha}a^{\dag}_{\ha}a^{\dag}_{\ha}$. Application of $\dot T_{{5\over
2}}^{(+)}$ forces the coefficient of the first term to vanish. Application of
$\dot T_{{3\over 2}}^{(+)}$ forces the coefficients of the next two terms to
vanish, and application of $\dot T_{\ha}^{(+)}$ forces the the coefficients
of the last two terms to vanish. Hence there are no physical states at
level three in the pure $\s$ theory.

At level four there are ten possible operators, namely
$$
a^{\dag}_2,\
b^{\dag}_1,\ a^{\dag}_{{3 \over 2}}a^{\dag}_{\ha},\
a^{\dag}_1a^{\dag}_1,\ a^{\dag}_1b^{\dag}_0,\
b^{\dag}_{\ha}a^{\dag}_{\ha},\ a^{\dag}_1a^{\dag}_{\ha}a^{\dag}_{\ha},\
b^{\dag}_0b^{\dag}_0,\
b^{\dag}_0a^{\dag}_{\ha}a^{\dag}_{\ha}, {\rm and}\
a^{\dag}_{\ha}a^{\dag}_{\ha}a^{\dag}_{\ha}a^{\dag}_{\ha}\ .
$$
Applying $\dot T_{{7\over 2}}^{(+)}$ to the general level $4$ state forces
the coefficient of the first term to vanish. $\dot T_{{5\over 2}}^{(+)}$
eliminates the next two terms, $\dot T_{{3\over 2}}^{(+)}$ eliminates the
fourth and fifth operators, while $\dot T_{2}^{(+)}$ and $\dot T_{1}^{(+)}$
eliminate the sixth and seventh, respectively. We are left then with products
of
the same operators $b^{\dag}_0$ and $a^{\dag}_{\ha}$ which survived at level
two. Applying the last constraint $\dot T_{\ha}^{(+)}$ forces these operators
to
appear in precisely the same combination found at the lower level, squared,
that
is the allowed state at level four is of the form,
$$
\vert 4, p\rangle=\left(\a_{\mo\mt} a^{\dag}_{\ha \mo}a_{\ha
\mt} +
\b b^{\dag}_{000}\right)^2\vert 0, p\rangle\ ,
\equn
$$
where the same conditions (i) or (ii) of {\conds} must be satisfied.

This procedure can be continued indefinitely. At each level $N$, applying the
$\dT$ constraints in the decreasing order $J= N -\ha, \dots, \ha$ eliminates
all
the possible linear combinations of operators at that level except for $N$
even,
where just that combination which appears at level $2$  survives, raised to the
power $N/2$.

It is not difficult to check directly that the state
$$
\vert 2n, p\rangle = \left(\sum_{\mo\mt}\a_{\ha\mo\mt} a^{\dag}_{\ha \mo}a_{\ha
\mt} + \b b^{\dag}_{000} \right)^n \vert 0, p\rangle
\equn\put\Nphys
$$
indeed does satisfy all the volume non-preserving diffeomorphism constraints,
$$
\dT \vert 2n, p\rangle = 0
\equn
$$
at general even level $N=2n$, provided (i) or (ii) of {\conds} is satisfied.
Taking careful note of the explicit form of $\dT$ {\mons} with the restrictions
on $J_1$ and $J_2$ in the sum, one quickly observes that most of the bilinear
operators annihilate the state  $\vert 2n, p\rangle$ and therefore give no
condition.  Explicitly, the $a_{_{\jo\mo}}a_{_{\jt\mt}}$ terms annihilate the
state unless $\jo = \jt = \ha$. The angular momentum coupling coefficient then
vanishes unless $J=0$ or $J=1$, but in either of those cases the coefficient
in
curly brackets vanishes in {\mons} for this term. The $a_{_{\jo\mo}}
b_{_{\jt\mt}}$ terms do not contribute unless $\jo =\ha$ and $\jt =0$ which
implies $J=\ha$ but in this case the corresponding coefficient again vanishes.
The $b_{_{\jo\mo}}b_{_{\jt\mt}}$ do not contribute since $\jo =\jt = 0$
implies $J=0$ and the corresponding coefficient in curly brackets vanishes. The
${\tilde a}^{\dag}_{_{\jt\mt}}a_{_{\jo\mo}}$ terms do not contribute unless
$\jo
=\ha$ but then $\jt$ cannot satisfy the restriction $\ha >\jt >0$. The
${\tilde b}^{\dag}_{_{\jt\mt}} a_{_{\jo\mo}}$ terms do not contribute
unless $\jo = \ha$ but then $\jt$ cannot satisfy the restriction
$\ha > \jt + 1$. The ${\tilde a}^{\dag}_{_{\jt\mt}} b_{_{\jo\mo}}$
terms do not contribute unless $\jo = 0$ but then $\jt = \ha$ and
only the $J=\ha$ moment must be examined in detail. Finally,
the ${\tilde b}^{\dag}_{_{\jt\mt}}  b_{_{\jo\mo}}$ terms do not contribute
unless $\jo =0$ but the restriction $0 > \jt$ cannot be satisfied. Since
from the linear terms in {\mons} only the $a_{_{\ha M}}$ term can contribute
we have explicitly the only surviving terms at $J=\ha$,
$$
\dot T^{(+)}_{\ha M} \vert 2n, p\rangle = -\left ((Q-ip)a_{_{\ha M}} +
{\tilde a}^{\dag}_{_{\ha M}} b_{00}\right) \vert 2n, p\rangle\ .
\equn
$$
Now one can verify that the commutator
$$
\left[ (Q-ip)a_{_{\ha M}} + {\tilde a}^{\dag}_{_{\ha M}} b_{00}\ ,
\a_{\ha\mo\mt} a^{\dag}_{\ha \mo}a^{\dag}_{\ha \mt} + \b b^{\dag}_{00}
\right] = 0
$$
if and only if one of the conditions {\conds} are satisfied. In that case the
surviving operator above passes through the operator creating the state in
{\Nphys} and annihilates the Fock vacuum. Hence the state $\vert 2n, p\rangle$
in {\Nphys} satisfies all the volume non-preserving spatial diffeomorphism
constraints if the conditions {\conds} are satisfied.

By a similar exercise it is not difficult to check that the state $\vert 2n,
p\rangle$ satisfies all the constraints of volume preserving spatial
diffeomorphisms as well, {\it viz.}
$$
X^{(+)}_{J \cm} \vert 2n, p\rangle = 0\ .
\equn
$$
Indeed, by inspecting the explicit form {\monspa} for $X^{(+)}_{J \cm}$
one quickly verifies that operating on this state the
$a_{_{\jo\mo}}a_{_{\jt\mt}}$ terms do not contribute since $\jo =\jt =\ha$ has
vanishing $\jo -\jt$. The $a_{_{\jo\mo}}b_{_{\jt\mt}}$,
$b_{_{\jo\mo}}b_{_{\jt\mt}}$, ${\tilde a}^{\dag}_{_{\jt\mt}} b_{_{\jo\mo}}$ and
${\tilde b}^{\dag}_{_{\jt\mt}} b_{_{\jo\mo}}$ terms do not contribute unless
$\jt = 0$ which is not in the primed sum. Likewise, the
${\tilde b}^{\dag}_{_{\jt\mt}} a_{_{\jo\mo}}$ term does not contribute
unless $\jo = \ha$ but then the restriction $\ha \ge \jt + 1$
excludes this term from the sum. Finally only the
${\tilde a}^{\dag}_{_{\jt\mt}}a_{_{\jo\mo}}$ term remains with $\jo = \ha$.
But then $\jo \ge \jt >0$ implies $\jt = \ha$ as well. Since in that case
$$
\r^i_{_{M_1M_2}} = i{V\over 4}\left( Y^*_{{1\over 2} M_1}\de^i\, Y_{{1\over 2}
M_2} - Y_{{1\over 2} M_2}\de^i\,Y^*_{{1\over 2} M_1}\right)  = \sqrt{2\over
V}\,\cy^i_{\ha\cm}
\equn\put\ks
$$
is just the lowest of transverse vector harmonics which are mutually orthogonal
on $S^3$, the coupling coefficient $\ck^{J\cm}_{\ha\mo\ha\mt}$ vanishes unless
$J=\ha$, so that only the $X^{(+)}_{\ha \cm} \sim R_{_{\mo\mt}}$ moment
condition needs to be verified in detail.  Using the properties of $\r^i$
given in the Appendix of paper I, this condition becomes
$$
R_{_{\mo\mt}}\vert 2n, p\rangle = \left(a^{\dag}_{_{\ha\mt}}a{_{\ha\mo}}
- \e_{_{\mo}}\e_{_{\mt}}a^{\dag}_{_{\ha -\mo}}a{_{\ha -\mt}}\right)
\vert 2n, p\rangle = 0\ ,
\equn
$$
which is satisfied if and only if the commutator
$$
\left[ a^{\dag}_{_{\ha\mt}}a{_{\ha\mo}}
- \e_{_{\mo}}\e_{_{\mt}}a^{\dag}_{_{\ha -\mo}}a{_{\ha -\mt}},
\sum_{M_3M_4}\a_{\ha M_3M_4} a^{\dag}_{\ha M_3}a^{\dag}_{\ha M_4}
\right] = 0\ ,
$$
which is satisfied in turn if and only if
$$
\a_{\ha M_1M_2} = \a \e_{_{\mo}}\d_{_{\mo\,-\mt}}
\equn\put\rotinv
$$
for some $\a$. This is just the condition that the state be $O(4)$
rotationally invariant. Recalling now the conditions {\conds} one finds that
in both cases (i) and (ii) we may identify $\b = 2(Q-ip)\a$ and
{\rotinv} is thereby satisfied. Hence, the state
$$
\vert 2n, p \rangle = \left(2(Q-ip)b_{00}^{\dag} +
\sum_M \e_{_M}a^{\dag}_{\ha M}a^{\dag}_{\ha -M}\right)^n
\vert 0, p \rangle
\equn\put\phst
$$
satisfies this constraint of rotational invariance, and all the spatial
diffeomorphism constraints.

The first remark to be made about this result is that it is interesting
that {\it any} non-trivial physical states survive the application of the
diffeomorphism constraints in the pure $\s$ theory. In two dimensions the
corresponding procedure eliminates everything except the vacuum state.
Clearly, the situation in four dimensions is much richer, since we find a
physical state at every even level $N=2n$.

It is also worthwhile to remark that the imposition of the full $T^{00}$
constraints of the classical theory would eliminate {\it all} of the physical
states from the spectrum except the Fock vacuum. This is because $T^{00}$
contains a time-independent term for any $J$ proportional to
$$\eqalign{
&\sum_{j\mo\mt}\cc^{JM}_{j\mo j-\mt}\Bigl\{\left[-4j(2j+1) - \hbox{$2\over 3$}
J(J+1)(6j+1) + \hbox{$1\over 3j$}J^2(J+1)^2\right]a^{\dag}_{j\mo}a_{j-\mt}\cr
& + \left[4(j+1)(2j+1) - \hbox{$2\over 3$} J(J+1)(6j+5) +
\hbox{$1\over 3(j+1)$}J^2(J+1)^2\right]b^{\dag}_{j\mo}b_{j\mt}\cr
& + \left[-4(j+1)J(J+1) + \hbox{$1\over 3(j+1)$}J^2(J+1)^2\right]
\left(a^{\dag}_{j+1\mt}b_{j\mo} + b^{\dag}_{j\mo}a_{j+1\mt}\right) \Bigr\}.
\cr}
$$
Requiring this to vanish on the physical states would lead to the
trivial solution at level zero alone. However, we know that
imposing the $J=0$ moment of the naive $T^{00}$ constraint above
is flatly {\it inconsistent} with the quantum Hamiltonian constraint with
the correct subtraction determined in paper I.

Lastly, we consider the Hamiltonian constraint. From {\level} we see that the
Hamiltonian constraint determines the momentum eigenvalue $p_{2n}$ at every
even
level $2n$ to satisfy
$$
\ha p_{2n}^2 + \ha Q^2 + 2n = 4 \ .
\equn\put\LO
$$
Hence the Hamiltonian physical state condition {\LO} cannot be
satisfied for real $p$ if $Q^2 > 8$. This is similar to the situation in two
dimensions for central charge $c_m <1$. In string theory this fact is
interpreted to mean that $p$ is not actually a momentum at all but a timelike
quantity more analogous to energy in the target space. Then we may define
$$
p_{2n} \equiv iE_{2n} \equn
$$
and write the physical state condition in the form
$$
E_{2n} = +\sqrt{Q^2 - 8 + 4n} \, ,
\equn\put\ephy
$$
with $E_{2n}$ real and positive. Although this continuation results in
states which are not $\d$-function normalizable, we recall from $c_m\le 1$ in
2D
gravity that they are created by operators with well-defined scaling
properties.
In this region of central charge the theory has a metric interpretation since
these operators are matter primary fields dressed with normal ordered
exponentials of the conformal factor with real exponents. In the next section
we
will see how this interpretation is generalized in four dimensions.

Since we have regarded $Q^2$ as a free parameter let us consider the situation
when $Q^2 < 0$ which corresponds to $c_m > 25$ in non-critical string theory.
In
this region the Hamiltonian condition {\LO} can be satisfied with real momenta,
and the roles of the $a$ and $b$ oscillators are interchanged, the $a^{\dag}$
oscillators now creating negative norm states. The physical states are still of
the form {\phst} and manifestly have positive norm since only pairs of
$a^{\dag}$ operators appear together. This means that the negative norm states
have completely disappeared from the spectrum after applying the diffeomorphism
constraints. Therefore the pure $\s$ theory described by the fourth order
action
{\Seff} is unitary for $Q^2 <0$, and does not suffer from the problems plaguing
local higher derivative theories of gravity in four dimensions. Since the
addition of unitary matter and/or transverse graviton degrees of freedom should
not disturb this feature of the $\s$ theory, we expect that the elimination of
the negative norm states by the imposition of the diffeomorphism constraints
will persist in the presence of these fields. This is indeed the case for free
conformal scalar fields, as we verify in the end of the next section. The
proper
inclusion of transverse gravitons in this quantum $\s$ framework remains an
open
problem.

\vskip 2cm

\newcount\eqncount
\sectadv
\set\sect{4}
\beginsection{4. Operator Correspondence}

Since positive energy free matter fields give positive $Q^2$, which may well be
larger than $8$ [\cAMM], we return to this case and find the operators creating
the physical states {\phst} and their geometric interpretation.

By analogy with the $D=2$ case we might guess that the general physical state
of
the pure $\s$ theory at level $2n$ should correspond to dimension $4$ operators
of the form,
$$
\co^{(n)}_4 = :F_n [\de\s] e^{\b_n\s}: \equn\put\genop
$$
where $F_n$ is a polynomial containing derivatives up to order $2n$ in
$\s$ such that the full operator corresponds to a primary field
with definite conformal weight equal to $4$.
The conformal weight of a pure normal ordered exponential may be calculated
in the canonical oscillator framework from the commutator,
$$
i\left[ K^{(\pm)}_M, :e^{y\s}:\right] = \k^{(\pm)\,a}\de_a :e^{y\s}: +
{w(y)\over 4} :e^{y\s}: \de_a \k^{(\pm)\,a}\
\equn\put\cwex
$$
upon using the Baker-Campbell-Haussdorf formula and the commutator
{\pqc} for the zero mode,
$$\eqalign{
:e^{y\s}:&= e^{y \s^{(-)}} e^{y\s_0} e^{y\s^{(+)}}\cr
&= e^{y \s^{(-)}} e^{{\hat q}y\over Q} e^{y{\hat p}t\over Q}
e^{-iy^2t\over 2Q^2} e^{y\s_0} e^{y\s^{(+)}}, \cr}
\equn\put\noror
$$
where $\s^{(+)}$, $\s^{(-)}$, $\s_0$ are the positive, negative and zero
frequency components of $\s$. By inserting the mode expansions for
$\s$, into the exponentials and comparing frequency level by
frequency level with {\cwex}, we find that {\cwex} is satisfied with
$$
w(y) = y - {y^2 \over 2 Q^2}\ .
\equn
$$
The classical scaling dimension $y$ of the exponential is modified by a quantum
correction which agrees with the covariant one-loop Feynman diagram calculation
of ref. [\cAntMot].

Hence, the pure exponential operator
$$
\int \,d^4 x: e^{\b_0 \s}:
\equn\put\purexp
$$
is a diffeomorphism invariant scalar provided
$$
w(\b_0) = \b_0 - {\b_0^2 \over 2Q^2} = 4\ .
\equn\put\bcona
$$
The solution of this quadratic equation is
$$
\b_0 = Q(Q +ip_0) = Q(Q-E_0) \ ,
\equn\put\bconb
$$
with the same branch of the square root chosen as in {\ephy} in order to
reproduce the classical limit,
$$
\b_0 \rarr 4 \quad {\rm as} \quad Q^2 \rarr \infty\ .
$$
In this limit we see that the normal ordered exponential operator
with $\b_0$ given by {\bconb} is just the volume operator $\int d^4x\sqrt{-g}$
which is clearly diffeomorphism invariant.

By operating with this exponential on the Fock vacuum state
$\vert \O \rangle $ annihilated by $H$, $K_M^{(+)}$, and $R_{_{\mo\mt}}$
we obtain a state invariant under all spatial diffeomorphisms. Since
$$
H\vert \O \rangle = \ha (\hat p^2 + Q^2)\vert \O \rangle = 0
\equn
$$
the vacuum state $\vert \O \rangle$ must be an eigenstate of $\hat p$
with non-zero imaginary momentum. Choosing again the same sign of
the square root as in {\ephy},
$$
\hat p \vert \O \rangle = iQ \vert \O \rangle
\equn
$$
we find that the state constructed with the exponential operator
is an eigenstate of $\hat p$ with eigenvalue $p_0 = iE_0$,
$$\eqalign{
\hat p \int d^4x\, :e^{\b_0\s}: \vert \O\rangle &= \hat p
\int d^4x\, :e^{\b_0\s_0}: \vert \O\rangle\cr
&=\left[\hat p, e^{\b_0 \hat q \over Q}\right]
e^{\b_0 pt\over Q}e^{-i\b_0^2t\over 2Q^2}\vert \O\rangle +
\int d^4x\, :e^{\b_0\s}: \hat p \vert \O\rangle \cr
&= \left(-{i\b_0\over Q} +iQ \right) \int d^4x\, :e^{\b_0\s}:
\vert \O\rangle \cr
&= i\sqrt{Q^2 - 8} \int d^4x\, :e^{\b_0\s}: \vert y\rangle\ .\cr}
\equn\put\peig
$$
Finally, since the exponential operator has weight $4$ and the Hamiltonian on
$R\times S^3$ just measures the conformal weight [\cpapI], it follows that this
state created by the exponential also satisfies the physical state Hamiltonian
condition, $H\vert phys \rangle = 4\vert phys \rangle$. Combining this with the
eigenvalue of the momentum operator we conclude that the normal ordered
exponential acting on the vacuum $\vert \O \rangle$ creates the physical state
at level zero, {\it i.e.}
$$
\vert 0, p_0 \rangle \sim \int d^4x\, :e^{\b_0 \s}: \vert \O \rangle\ .
\equn\put\low
$$

To proceed with the operator identification at higher levels we need to
identify the proper primary operators of the form {\genop} that create the
higher level physical states. Since $F_n$ contains up to $2n$ derivatives of
the $\s$ field, we would like to find that the conformal weight of the
level $n$ operator be given by the formula,
$$
w_n = 2n + \b_n - {\b_n^2 \over 2 Q^2} = 4 \ ,\equn\put\bwei
$$
which would imply
$$
\b_n = Q(Q + ip_{2n}) = Q(Q-E_{2n}) \ , \equn
$$
in analogy with relation {\bconb} at level zero. However unlike in $D=2$ where
simple normal ordering is sufficient to remove all divergences, in $D=4$ normal
ordering operators of the form {\genop} only eliminates the leading (quartic)
divergences of these composite operators, leaving  behind, in general quadratic
divergences as well as the logarithms wanted  for conformal invariant behavior.
Hence, we cannot expect normal ordered  polynomial functions of $\s$ with a
single exponential to have well-defined  conformal dimensions or to create all
the physical states for arbitrary finite $Q^2$ in four dimensions. The operator
mixing problem requires  the power of the operator product expansion of field
theory and cannot be  analyzed easily in the Fock space of oscillators of the
present treatment. Indeed from the covariant $\s$ field theory analysis of
the $\b$-functions of the Einstein and cosmological couplings we know that
such operator mixing does occur in the $\s$ theory [\cAntMot]. Hence, we shall
have to content ourselves here with an identification of operators with the
physical states at $n>0$ only in the limit $Q^2 \rarr \infty$ where this
operator mixing problem disappears.

Since
$$
\b_n \rarr 4 -2n \quad {\rm as}\quad Q^2 \rarr \infty ,
\equn
$$
it is not difficult to find the functions $F_n$ explicitly in this limit.
Consider just the Ricci scalar of the full metric {\confdef} raised to the
power
$n$ multiplied by $\sqrt{-g}$, {\it i.e.}
$$\eqalign{
&:R^{n}\sqrt{-g}: =
:\left[ \sqb\s + \nabar^a\s\nabar_a\s -{\bR \over 6} \right]^n e^{(4-2n)\s}:\cr
&=:\left[-1 - \dot\s_o^2 + \sqb\s_q - 2\dot\s_0\dot\s_q
+ \nabar^a\s_q\nabar_a\s_q\right]^n e^{(4-2n)\s_0}e^{(4-2n)\s_q}:\cr
&=:\left[-1 - {p_{2n}^2 \over Q^2} - 2i (1-{ip_{2n}\over Q})\dot\s_2
+2i( 1+{ip_{2n}\over Q})\dot\s_1+\nabar^a\s_q\nabar_a\s_q  \right]^n
e^{(4-2n)\s_0}e^{(4-2n)\s_q}:\cr
&= :\left[-4i\dot\s_2 +\nabar^a\s_q\nabar_a\s_q + \dots\right]^n
e^{(4-2n)\s_0}e^{(4-2n)\s_q}:\cr
\cr}
\equn\put\levna
$$
where $\s_q = \s_1 + \s_2$ and the ellipsis contains terms with subleading
$Q^2$
dependence. In the large $Q^2$ limit $\s_q$ is held fixed while
$p_{2n}\rightarrow iQ$. Substituting the mode expansions for $\s_q$, using the
properties of the $J = \ha$ scalar harmonics catalogued in the Appendix of
paper
I, operating on the vacuum state, and integrating, one finds finally
$$
\int d^4\, x:\left[{4\over Q}e^{2it}b_{00}^{\dag} + {1 \over Q^2} e^{2it}
\sum_M \e_{_M}a^{\dag}_{\ha M}a^{\dag}_{\ha -M} + \dots \right]^n
e^{(4-2n)\s_0}:\vert \O\rangle\ .
\equn\put\newst
$$
Frequencies higher than $\exp (2it)$ have been dropped in this last ellipsis
since the time dependence of $\exp\left((4-2n)\s_0\right)$  acting on the
vacuum
$\vert \O \rangle$ becomes just $\exp (4-2n)it$  in the large $Q^2$ limit, so
that the total time dependence  of the terms exhibited becomes $\exp (4it)$. It
is this time dependence which is selected by the time integration as determined
by the conformal transformation from flat space described in paper I. Hence in
the classical $Q^2 \rarr \infty$ limit the only surviving operator structure in
{\newst} is just the linear combination {\phst}, and the physical state at
general level $2n$ corresponds to the integral of the Ricci scalar to the
$n^{th}$ power, {\it i.e}
$$
\vert 2n, p_n \rangle \rarr \int d^4x \, \sqrt {-g}\, R^n \ \vert \O \rangle
\equn
$$
as $Q^2 \rarr \infty$.

Notice that for $n=1$ this operator becomes just the integral of the Ricci
scalar in the large $Q^2$ limit. Moreover, for finite $Q^2$ the equation for
the
weight $\b_1$ in {\bwei} is identical to the equation for the scaling dimension
of the conformal factor which sets the weight of the Ricci scalar density equal
to 4,
$$
w(\sqrt{-g}R)\Big\vert_{g =  e^{2\a\s}\bar g} = 2\a - {2\a^2\over Q^2} + 2 = 4
\equn
$$
obtained by the covariant methods of [\cAntMot], with the identification
$\b_1 = 2\a$. This operator mixes with the pure exponential $\l\exp(4\a\s)$
in a well-defined way, the relative coefficient $\l$ determined by
eq. (3.17) of ref. [\cAntMot] and vanishing in the $Q^2 \rarr \infty$
classical limit. This mixing phenomenon at finite $Q^2$ is an example
of the operator mixing mentioned above which should generalize
to higher levels for the operators $\sqrt{-g} R^n$.

The pure exponential operator found at level zero {\purexp} with $\b_0$
determined by {\bcona} is identical to the volume operator added to the $\s$
action in ref. [\cSch]. This operator and all of the higher level operators
when
added to the action {\Seff} correspond  to conformal deformations of the
original ``free" $\s$ theory, as  encountered in two dimensional gravity with
matter. It is characteristic of the non-trivial dynamics of the conformal
factor
in $D=4$ that there are such diffeomorphic invariant operators, even without
the
addition of any matter or transverse graviton fields.

Finally, one can repeat the physical state analysis for a conformal scalar
field, $\Phi$, coupled to $\s$. The quantization and diffeomorphism generators
for the scalar field are given in paper I. The energy-momentum tensor is now
the
sum of the contributions from the $\Phi$ and $\s$ fields. Starting with the
Fock
vacuum for the combined matter plus $\s$ theory, the general level $N$ state
can
be constructed by acting with the creation operators $\varphi^{\dag}, a^{\dag},
b^{\dag}$. Applying first the volume non-preserving diffeomorphism constraints
$(\dT[\Phi] + \dT[\s]) \vert phys \rangle = 0$ leaves two kinds of states,
$$\eqalign{
\left(2(Q-ip)b_{00}^{\dag} +
\sum_M \e_{_M}a^{\dag}_{\ha M}a^{\dag}_{\ha -M}\right)^n
\left(\varphi_{00}^{\dag}\right)^l
&\vert 0, p \rangle \cr {\rm and}\qquad
\left(\sqrt{2}(Q-ip)\varphi_{\ha M}^{\dag} -
a^{\dag}_{\ha M}\varphi^{\dag}_{00}\right)
\left(\varphi_{00}^{\dag}\right)^l &\vert 0, p \rangle \ .\cr}
$$
However the second class of states are not rotationally invariant (as is
obvious
from the free magnetic index $M$) and are eliminated by the global rotation
constraint $R_{\mo\mt} \vert phys\rangle = 0$. Hence the surviving physical
states are a direct product of the states found in either the $\Phi$ or $\s$
theories separately. The Hamiltonian constraint at level $N = 2n + l$
determines
the momentum eigenvalue $p_N$ by
$$
\ha p^2_{2n , l} + \ha Q^2 + 2n + l = 4\ .
\equn
$$
Clearly, in the classical limit $Q^2 \rarr \infty $
the operators which create these physical states are of the form,
$$
\int d^4x \, \sqrt {-g}\, R^n \Phi^l\ \vert \O \rangle \ .
$$
This example shows that the results obtained for the physical states and
operators in the $\s$ theory are easily generalized in the presence of matter,
at least in this limit.

\vfill
\eject
\newcount\eqncount
\sectadv
\set\sect{5}
\beginsection{5. Summary and Discussion}

In paper I we studied the constraints of coordinate invariance for any quantum
theory possessing conformal symmetry independently of its particular dynamics.
The Dirac-Fock method of imposing the constraints requires only the vanishing
of
the matrix elements of the generators of spatial diffeomorphisms $\langle
T^{0i}\rangle$, because of their non-trivial commutators. We argued from the
structure of the quantum algebra off-shell that the time reparametrization
constraint generated by $T^{00}$ should not be imposed {\it a priori}, but
only the weaker condition $\langle \dot T^{00}\rangle = 0$. We found
also that the Hamiltonian which is the spatial integral of $T^{00}$
is modified by a finite and calculable shift of $-4$.

In this paper we applied this general formalism to the quantum theory of the
conformal factor generated by the trace anomaly in four dimensions. Despite the
fourth order effective action of this theory we found that the constraints of
invariance under spatial diffeomorphisms eliminate all negative norm states. In
the region of $Q^2 < 0$ normalizable states with real momentum exist and there
is no violation of unitarity. In the region $Q^2 > 8$ which we regard as the
physical region the physical states are non-normalizable and quite analogous to
the discrete states of 2D quantum gravity in the region $c_m \le 1$. The states
are created by invariant scalar operators which in the classical limit $Q^2
\rarr \infty$ correspond to integer powers of the scalar curvature integrated
over spacetime. The scaling dimensions of these operators in the canonical
approach provide an independent confirmation of the covariant treatment
initiated in ref. [\cAntMot]. Hence, the $\s$ effective action passes all the
tests of a sensible quantum theory.

In a sense it is not surprising that the constraints of diffeomorphism
invariance eliminate the negative norm states since this occurs even in the
linearized Einstein theory. What is more remarkable is that there are any
non-trivial states at all which survive in the complete absence of transverse
gravitons. This confirms that the conformal factor is indeed dynamical (though
non-propagating) in the full quantum theory, quite independently of the
graviton
degrees of freedom. Moreover,  the physical states in this theory are created
by
operators possessing clear geometric significance in terms of the scalar
curvature which is completely fixed by the equations of motion of the Einstein
theory, but which fluctuates in the quantum theory. This is in contrast to two
dimensions where the only effect of the Liouville mode is the gravitational
dressing of matter operators.

In additional to some technical problems such as a more precise
characterization
of the operator mixing problem at finite $Q^2$, several issues remain to be
addressed. Among these is the inclusion of transverse graviton modes, which
although not expected to change the framework drastically should be
investigated
carefully. Another open problem is the correct characterization of the full
algebra of quantum diffeomorphisms off-shell, which seems to be required for
invariance under finite diffeomorphisms. If such a general algebra does not
exist there may be serious implications for the consistency of the theory
and/or
for the Dirac-Fock approach to quantization which we have followed. Finally,
the
calculation of correlation functions in the $\s$ theory should provide
non-trivial information about the far infrared behavior of quantum gravity in
four dimensions with observable consequences in cosmology and the large scale
structure of th Universe.

\vskip 1.5truecm
\centerline{{\it Acknowledgments}}
\vskip .5cm

I.A. and P.O.M. would like to acknowledge the hospitality of
Theoretical Division (T-8) of Los Alamos National Laboratory.
E.M. and P.O.M. would like to thank the Centre de Physique Th\'eorique
of the Ecole Polytechnique for its hospitality. All three authors
wish to acknowledge NATO grant CRG 900636 for partial financial support.
\vfill
\eject

\baselineskip=15pt
{\bf REFERENCES }
\vskip.5cm

\item{[\cAntMot]} I. Antoniadis and E. Mottola, {\it Phys. Rev.} {\bf D45}
(1992) 2013; \hfill\break\noindent I. Antoniadis, P. O. Mazur and E. Mottola,
{\it Phys. Lett.} {\bf B323} (1994) 284.\hfill\break

\item{[\cAMM]} I. Antoniadis, P. O. Mazur, and E. Mottola, {\it Nucl.
Phys.} {\bf B 388} (1992) 627.\hfill\break

\item{[\cPoly]} A. M. Polyakov, {\it Phys. Lett.} {\bf B103} (1981) 207, 211;
{\it Gauge Fields and Strings}, \hfill\break\indent
(Harwood Academic, Chur, 1987).\hfill\break

\item{[\cdeS]} B. Allen and G. Turyn, {\it Nucl. Phys.} {\bf B292} (1987)
813;\hfill\break E. G. Floratos, J. Iliopoulos and T. N. Tomaras, {\it Phys.
Lett.} {\bf B197} (1987) 373;\hfill\break I. Antoniadis and E. Mottola, {\it
Jour. Math. Phys.} {\bf 32} (1991) 1037.\hfill\break

\item{[\cKPZ]} V. G. Knizhnik, A. M. Polyakov, and A. B. Zamolodchikov,
{\it Mod. Phys. Lett.} \hfill\break\indent{\bf A3} (1988) 819.\hfill\break

\item{[\cYork]} J. W. York, {\it Phys. Rev. Lett.} {\bf 26} (1971) 1656;
{\it ibid.} {\bf 28} (1972) 1082; \hfill\break\indent
{\it Jour. Math. Phys.} {\bf 13} (1972) 125;
{\it ibid.} {\bf 14} (1973) 125.\hfill\break

\item{[\cpapI]} I. Antoniadis, P. O. Mazur and E. Mottola, Ecole Polytechnique
preprint \hfill\break\indent CPTH-S376.0995, and Los Alamos preprint
LA-UR-95-3392, \hfill\break\indent
referred to as paper I in this work.\hfill\break

\item{[\cLiou]} T. L. Curtright and C. B. Thorn, {\it Phys. Rev. Lett.} {\bf
48}
(1982) 1309; \hfill\break\indent {\it Erratum, ibid.}
1768;\hfill\break\noindent
J. L. Gervais and A. Neveu, {\it Nucl. Phys.} {\bf B199} (1982) 59;
\hfill\break\indent {\bf B209} (1982) 125; {\bf B224} (1983) 329.\hfill\break

\item{[\cLBC]} L. S. Brown and J. P. Cassidy, {\it Phys. Rev.} {\bf D16} (1977)
1712.\hfill\break

\item{[\cGSW]} M. B. Green, J. H. Schwarz, and E. Witten, {\it Superstring
Theory, Vol. 1}, \hfill\break\indent (Cambridge Univ. Press, Cambridge,
1987).\hfill\break

\item{[\cSch]} C. Schmidhuber, {\it Nucl. Phys.} {\bf B390} (1993) 188.

\vfill
\eject

\end